\documentclass[aps,prd,twocolumn,groupedaddress,showpacs]{revtex4-1}

\usepackage{graphicx}
\usepackage{dcolumn}
\usepackage{bm}

\begin{document}

\title{Constraining the photon flux in two-photon processes at the LHC}

\author{G. G. da Silveira}

\email[]{gustavo.silveira@cern.ch}

\affiliation{High and Medium Energy Group \\ Instituto de F\'{\i}sica e Matem\'atica, Universidade Federal de Pelotas\\
Caixa Postal 354, CEP 96010-090, Pelotas, RS, Brazil}

\author{V. P. Gon\c calves}

\email[]{barros@ufpel.edu.br}

\affiliation{High and Medium Energy Group \\ Instituto de F\'{\i}sica e Matem\'atica, Universidade Federal de Pelotas\\
Caixa Postal 354, CEP 96010-090, Pelotas, RS, Brazil}

\begin{abstract}
In this paper we propose the study of the $W^+W^-$ and $\mu^+\mu^-$ production by two-photon interactions in $pp$ collisions at LHC energies  to constrain the photon flux associated to an ultrarelativistic proton.  We consider the current parametrizations for the photon distribution of the proton and estimate the effective photon - photon luminosities for elastic, semielastic and inelastic processes. Moreover, we present predictions for the rapidity and invariant mass distributions for the two-photon
production of $W^+W^-$ and $\mu^+\mu^-$ in $\gamma \gamma$  interactions at LHC energies. We demonstrate that the semielastic and inelastic predictions are strongly dependent on the description of the inelastic photon flux, and that the relative contribution of the different processes depends on the invariant mass of the final state. Our results implies that   a dedicated experimental analysis of the  two-photon production of $W^+W^-$ and $\mu^+\mu^-$ with the tagging of one of the  protons in the final state  can be useful to constrain the magnitude of the inelastic component of the photon distribution.
\end{abstract}

\pacs{xxx}

\maketitle

\section{Introduction}

Ultrarelativistic charged hadrons (protons or nuclei)  give rise to strong electromagnetic fields. 
Consequently, in hadronic collisions at the  LHC,  the photon stemming from the electromagnetic field of one of the two colliding hadrons can interact with one photon of the other hadron  or can interact directly with the other hadron  \cite{upc}. The study of these photon - photon and photon - hadron interactions offers a unique opportunity to study fundamental aspects of Quantum Electrodynamics (QED) and Quantum Chromodynamics (QCD). In particular, the two-photon particle production have been extensively discussed in literature (For reviews see, e.g., Refs.  \cite{epa,Terazawa,Baurjpg,Baurdileptons,Chernyak}), with a recent revival of interest due to the experimental 
results on the exclusive two-photon production of $W^+W^-$ and $\ell^+\ell^-$ pairs by $\gamma\gamma$ interactions reported by the CMS Collaboration \cite{Chatrchyan:2011ci,Chatrchyan:2012tv,Chatrchyan:2013foa}. The observation of these processes demonstrated the feasibility of measuring such events within the current experimental apparatus available at the LHC, allowing novel studies of QCD at very high energies and searches for Beyond Standard Model Physics (See, e.g., Ref. \cite{newphysics}).

The basic ingredient in the analysis of the photon - induced processes is the description of the  equivalent photon distribution of the hadron, given by $\gamma(x,\mu^2)$, where $x$ is the fraction of the hadron energy carried by the photon and $\mu$ has to be identified with a momentum scale of the  process. The equivalent photon approximation of a charged  pointlike fermion was formulated  many years ago by Fermi \cite{Fermi} and developed by Williams \cite{Williams} and Weizsacker \cite{Weizsacker}. In contrast, the calculation of the photon distribution of the hadrons still is a subject of debate, due to the fact that they are not pointlike particles. In this case it is necessary to distinguish between the  elastic and inelastic components.  The elastic component, $\gamma_{el}$, can be estimated analysing the transition $h \rightarrow \gamma h$ taking into account the effects of the hadronic form factors, with the hadron remaining intact in the final state \cite{epa,kniehl}. In contrast, the inelastic contribution, $\gamma_{inel}$, is associated to the transition $h \rightarrow \gamma X$, with $X \neq h$, and  can be estimated taking into account the partonic structure of the hadrons, which can be a source of photons (See, e.g. Refs. \cite{epa,rujula,drees_godbole,pisano,mrstqed,nnpdf,cteqqed,martin_ryskin}). 

In a recent paper \cite{vicgus} we have proposed, for the first time, the study of the diffractive quarkonium photoproduction in $pp$ collisions at  LHC energies as a probe of the photon distribution of the proton. Our results indicated that, for the models considered, the contribution of the inelastic processes is of the same order or larger than the elastic one, with the predictions for the rapidity distributions being largely different, which makes the experimental discrimination feasible, with the detection of the two protons into the final state being indispensable to separate the inelastic and elastic events. Our goal in this paper is to extend our previous analysis for the  $W^+W^-$ and $\mu^+\mu^-$ production by two-photon interactions  in $pp$ collisions at LHC energies (For related studies see Refs. \cite{antoni_gus,antoni_royon}). Our study is strongly motivated by the recent results reported by the CMS Collaboration, which shows that the experimental technique currently applied allows the measurement of the two-photon production of pairs with large invariant mass in a scenario with non-negligible pileup \cite{Chatrchyan:2013foa}, and by the perspective of setup of forward detectors by the CMS and TOTEM Collaborations \cite{Albrow:2008pn,ctpps}, which will enhance the kinematic coverage for such investigations, bringing more data to elucidate the proper model for the photon. It is important to emphasize that the uncertainty present on the current measurements is an important aspect regarding the observation of rare processes that have not been measured in laboratory, like the exclusive production of photon pairs in $pp$, $p$Pb and PbPb collisions, with predictions presented for the first time in Ref.~\cite{d'Enterria:2013yra}. Then, the study of the different approaches for the photon flux is important to provide accurate predictions to be compared with data. In this sense, the purpose of this paper is to investigate the available approaches for the two-photon processes in $pp$ collisions and  to determine  the regions of the phase space that allow the discrimination among these possible models.

This paper is organized as follows. In the next Section we present a brief review of the formalism for the two-photon particle production and discuss the different models for the photon flux. In Section \ref{sec:res}, we present our results for the effective photon - photon luminosities and for the two-photon $W^+W^-$ and $\mu^+\mu^-$ differential cross sections. Finally, in Section~\ref{sec:sum} we summarize our main conclusions.

\begin{widetext}

\begin{figure}[t!]
\includegraphics[scale=0.45]{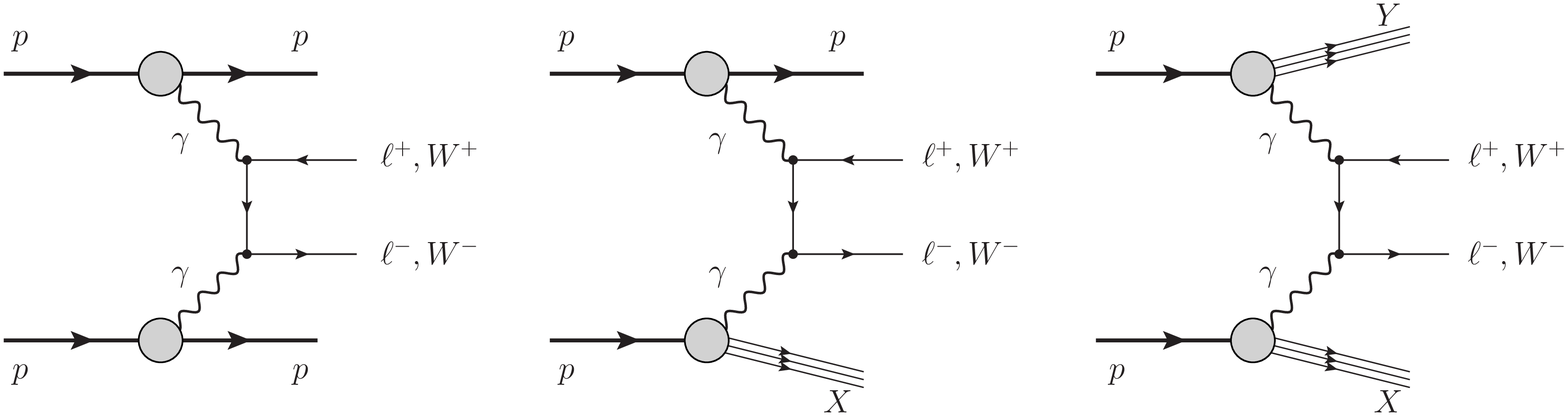}
\caption{Two-photon particle production in elastic (left), semielastic (center) and inelastic (right) processes.}
\label{diagrams}
\end{figure}

\end{widetext}

\section{Two-photon particle production}
\label{sec:form}

Following Ref.~\cite{KMR} we will write the cross sections for the  $\gamma \gamma$ production of a final state $F$ $ (= W^+W^-$ or $\mu^+\mu^-)$ of invariant mass $M=W_{\gamma\gamma}$ in a factorized form:
\begin{eqnarray}
\sigma = {\cal{L}}_{eff}(M^2,Y) \hat{\sigma}(M^2)
\label{fac}
\end{eqnarray}
where $\hat{\sigma}$ is the cross section for the hard subprocess $\gamma\gamma\rightarrow F$ and ${\cal{L}}_{eff}$ is the effective photon - photon luminosity for the production of the  system $F$ at rapidity $Y$. The effective luminosity is given in terms of the photon distribution of the incident hadrons as follows
\begin{eqnarray}
\frac{\partial {\cal{L}}_{eff}}{\partial Y \partial \ln M^2} = x_1\gamma(x_1,\mu^2) \cdot x_2\gamma(x_2,\mu^2) \,\,.
\label{lum1} 
\end{eqnarray}
The photon distribution of a nucleon consist of two parts: the elastic and inelastic components. In the elastic case, we have the coherent emission of photons from the hadron, without the dissociation of the incident hadron.
In contrast, in the inelastic case the photons are emitted by the quarks and antiquarks present in the hadrons, and the incident hadron is excited in a low-mass state. Consequently, we can define three different classes of $\gamma \gamma$ events: (a) the {\it elastic} processes, where the final state $F$ is produced with the two incident hadrons remaining intact, and the effective luminosity is proportional to  $ x_1\gamma^{el}(x_1,\mu^2) \cdot  x_2 \gamma^{el}(x_2,\mu^2) $; (b) the {\it semielastic}   processes, where one of the incident hadrons remain intact and the other dissociates, with $
{\cal{L}}_{eff} \propto [x_1\gamma^{el}(x_1,\mu^2) \cdot  x_2\gamma^{inel}(x_2,\mu^2) + x_1\gamma^{inel}(x_1,\mu^2) \cdot  x_2\gamma^{el}(x_2,\mu^2)]$
and (c) {\it inelastic} processes, where the two incident hadrons dissociates and
 $ {\cal{L}}_{eff} \propto x_1\gamma^{inel}(x_1,\mu^2) \cdot  x_2\gamma^{inel}(x_2,\mu^2)$. These three classes are represented in Fig.~\ref{diagrams}. In all these processes, the final state will be characterized by the presence of the state $F$ and two rapidity gaps, with the hadrons or the low-mass hadron beam fragments traveling in the beam direction (For related studies see, e.g. Ref. \cite{inel}).
 
In principle, these different processes can be separated by the tagging of the two very forward scattered hadrons and by the requirement of the presence of large rapidities gaps in the central detector. Unfortunately, forward detectors were not available during Run I of the LHC, and the next run starting in 2015 is going to produce a sizable pile-up obliterating the observation of rapidity gaps. Therefore, experimental separation of those contributions is a hard task and demands better knowledge of final-state kinematics in an observed event. In Run I, the CMS Collaboration have separated the signal of two-photon production of pairs ($W^+W^-$ and $\mu^+\mu^-$) by selecting lepton tracks from the information recorded in the tracking system, which can be used to analyze exclusive events even in a scenario with large number of interaction per bunch crossing (high pileup). Offline, events have been selected with no additional tracks associated to the $\ell^{+}\ell^{-}$ vertex. Since both elastic and non-elastic processes contribute in this case, we have higher photon luminosities, although larger uncertainties in the theoretical predictions, due to the lesser theoretically controlled inelastic photon flux (see below).  
Currently, there is a great expectation due to the installation of the CMS-TOTEM Precision Proton Spectrometer (CT-PPS) \cite{ctpps}, which will be setup in a first stage in one of the CMS sides at about 200 metres from the interaction point. Certainly, this new detector will improve the analysis of exclusive processes and allow us to access a variety of physics topics at high luminosities. However, a precise determination of the semielastic contribution will be still fundamental before the search of New Physics in two-photon processes. 

\begin{widetext}

\begin{figure}[t]
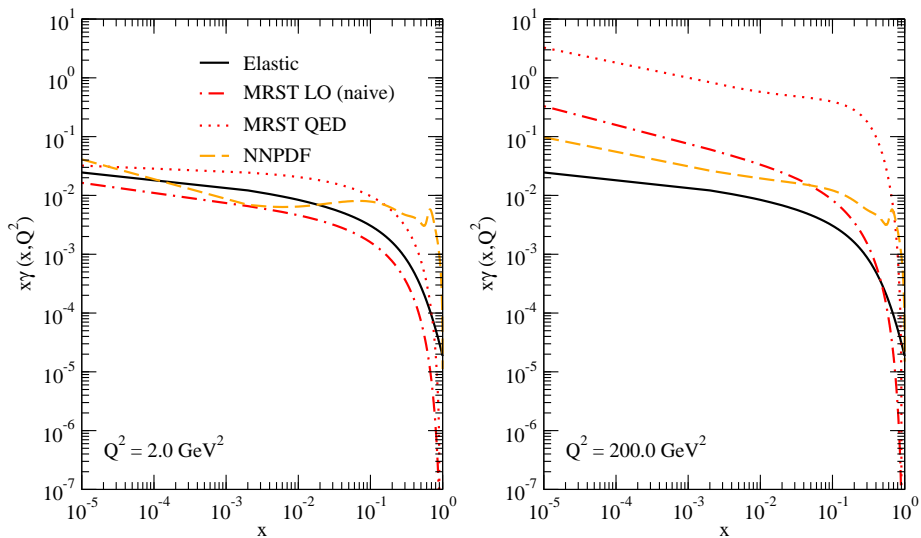

\begin{tabular}{cc}
 \includegraphics[scale=0.45]{densities_Mgaga_sqrt2GeV2_clean_x10m5.eps} &
 \includegraphics[scale=0.45]{densities_Mgaga_sqrt200GeV2_clean_x10m5.eps}
 \end{tabular}
 \caption{\label{densities_inel}
(Color online) Comparison between the different models for the inelastic photon distribution for two different values of the hard scale  $\mu^2 = Q^2$. The elastic component is presented for comparison.}
\end{figure}

\end{widetext}
 
We have that the effective luminosities are directly related to the elastic and inelastic photon distributions.  In order to compare our results for the elastic effective luminosities with those presented in Ref. \cite{KMR}, in what follows we will assume that the elastic photon distribution is given by approximated expression
 \begin{eqnarray}
 \gamma^{el} (x) = \frac{\alpha_{em}}{\pi} \frac{1}{x} \int_{\kappa_{min}^2} d\kappa^2 \frac{\kappa^2-\kappa_{min}^2}{\kappa^4} |F(\kappa^2)|^2 \,\,,
  \label{kmr_elastic} 
 \end{eqnarray}
 where $\kappa^2$ is the momentum transfer from the projectile, which has a form factor $F(\kappa^2)$, $\kappa_{min}^2 = (x\,m_N)^2$ and $m_N$ is the nucleon mass. 
A comparison between the different models for the elastic photon distribution is presented in Ref. \cite{vicwerdaniel}.  
In the last years, the inelastic  photon distribution of a nucleon has been derived considering different assumptions. In Ref.~\cite{drees_godbole}, a naive approach to the photon flux has been proposed, with the photon distribution in the proton given by a convolution of the distribution of quarks in the proton and the distribution of photons in the quarks as follows
\begin{eqnarray} \nonumber
\gamma^{inel}(x,\mu^ 2) = \sum_q \int_x^1 \frac{dx_q}{x_q} f_q(x_q,\mu^2) e_q^2 f_{\gamma/q}\left(\frac{x}{x_q},Q_1^2,Q_2^2\right), \\ \label{naive}
\end{eqnarray}   
where the sum runs over all quark and antiquark flavours and the flux of photons in a quark $f_{\gamma/q}$ is given by
\begin{eqnarray}
 f_{\gamma/q} (z) = \frac{\alpha}{2\pi} \frac{1+(1-z)^2}{2} \log \frac{Q_1^2}{Q_2^2} \,\,,
\end{eqnarray}  
with $Q_1^2$ being assumed to be the maximum value of the momentum transfer in the process and $Q_2^2$ is assumed to be equal to 1~GeV$^2$ in order to the parton model to be applicable.  On the other hand, different groups have studied the modification of  the Dokshitzer-Gribov-Lipatov-Altarelli-Parisi (DGLAP) equations for the quark and gluon distributions by the inclusion of QED contributions and have performed global parton analysis of deep inelastic and related hard-scattering data \cite{mrstqed,nnpdf,cteqqed,martin_ryskin}. 
Basically, the DGLAP equations and the momentum sum  rule are modified considering the presence of the  photon as an additional pointlike parton in the nucleon. The parametrizations for the photon distribution currently available in the literature \cite{mrstqed,nnpdf} differ in the approach for the initial condition for the photon distribution. While the Martin-Ryskin-Stirling-Thorne~(MRST) group assume that  $\gamma^{inel} (x,Q_0^2)$ is given by a expression similar to Eq.~(\ref{naive}), the Neural Network Parton Distribution Functions~(NNPDF) group parametrize the  input photon PDF and attempt to determine the parameters from the global data. The preliminary analysis by the Coordinated Theoretical-Experimental Project on QCD~(CTEQ) group, presented in Ref.~\cite{cteqqed}, assume a similar theoretical form for  $\gamma^{inel} (x,Q_0^2)$ to that proposed by the MRST group, but with an arbitrary normalization parameter, which is expressed as the momentum fraction carried by the photon. More recently, a distinct approach for the initial condition for the evolution of the photon distribution has been introduced in Ref.~\cite{martin_ryskin}, where the authors have proposed  that the starting distribution for the photon PDF
should be the total photon distribution, i.e., by the sum of the elastic and inelastic components. The main motivation of this approach is the reduced uncertainty in the input photon PDF, since the major part of the distribution is given by the elastic component, which is well  known. As a consequence of this assumption, the elastic component is dominant also at large values of the hard scale $\mu^2$ (See Fig.~5 in \cite{martin_ryskin}).
Unfortunately, the current data is not sufficient accurate to precisely determine the initial condition. As a consequence, the current predictions for the inelastic photon component strongly differ in its $x$ dependence. In what follows we  will consider  the MRST2004QED and NNPDF parametrizations, since only these two are currently available to  public use. 
In Fig.~\ref{densities_inel} we present the predictions of the MRST2004QED and NNPDF parametrizations for the inelastic photon distribution considering two different values for the hard scale $\mu^2$. For comparison the predictions of the naive approach [Eq.~(\ref{naive})] and the elastic component [Eq.~(\ref{kmr_elastic})] are also presented. While the elastic component is independent of the hard scale $\mu^2$, the inelastic component is strongly dependent, increasing at larger values of $\mu^2$. Moreover, all models predict that the inelastic contribution is dominant at very small values of $x$ and large $\mu^2$. However, as demonstrated in the figure, the $x$-dependence of the inelastic parametrizations is very distinct. 
 
In  order to estimate the two-photon $W^+W^-$ and $\mu^+\mu^-$ production  in $pp$ collisions we need to specify the cross section for the hard subprocess $\gamma \gamma \rightarrow a^{+}a^{-}$ ($a=W,\mu$). In what follows we will assume for the production of muons pairs
\begin{widetext}
\begin{eqnarray}\nonumber
\hat{\sigma}_{\gamma\gamma\to\mu^{+}\mu^{-}}(M^{2}) = \frac{4\pi\alpha_{em}^{2}}{M^{2}} \left\{ 2 \ln \left[ \frac{M^{2}}{2m_{\mu}}\left(1 + \sqrt{1 - \frac{4m_{\mu}^{2}}{M^{2}}}\right)\right]\left( 1 + \frac{4m_{\mu}^{2}M^{2}-8m_{\mu}^{4}}{M^{4}} \right) - \left( 1+ \frac{4m_{\mu}^{2}M^{2}}{M^{4}} \right)\sqrt{1 - \frac{4m_{\mu}^{2}}{M^{2}}}\right\}, \\
\end{eqnarray}
\end{widetext}
called the Breit-Wheeler formula, and for the production of $W$ pairs we employ \cite{Boonekamp:2007iu}
\begin{widetext}
\begin{eqnarray}\nonumber
\hat{\sigma}_{\gamma\gamma\to W^{+}W^{-}}(M^2) = \frac{8\pi\alpha_{em}^{2}}{M^{2}} \left[ \left( \frac{m^{2}_{W}}{M^{2}} + \frac{3}{4} + \frac{3M^{2}}{m^{2}_{W}} \right) \sqrt{1-\frac{4M^{2}}{m^{2}_{W}}} - \frac{3M^{2}}{m^{2}_{W}}\left(1-\frac{M^{2}}{m^{2}_{W}}\right)\log\left( \frac{1+\sqrt{1-4M^{2}/m^{2}_{W}}}{1-\sqrt{1-4M^{2}/m^{2}_{W}}} \right) \right]. \\
\end{eqnarray}
\end{widetext}

\begin{widetext}

\begin{figure}[t]
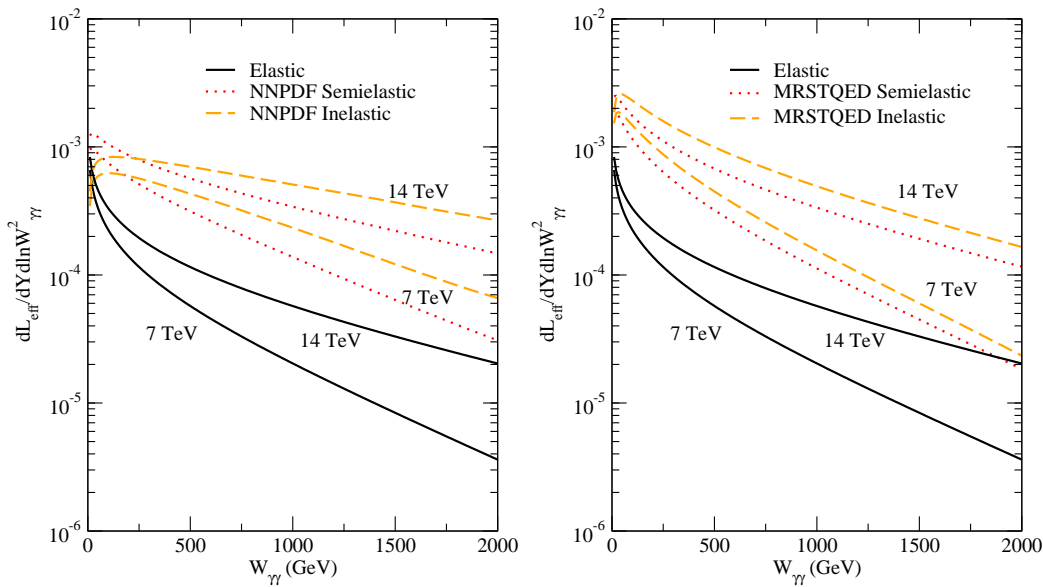

\begin{tabular}{cc}
\includegraphics[scale=0.45]{lumi_invmass_NNPDF_clean.eps} &
\includegraphics[scale=0.45]{lumi_invmass_MRSTQED_clean.eps}
\end{tabular}
\caption{\label{eff_lumi_naive} (Color online)
Effective photon - photon luminosity for elastic, semielastic and inelastic processes for two different center-of-mass energies considering the MRSTQED and NNPDF parametrizations.
}
\end{figure}

\end{widetext}

\section{Results}
\label{sec:res}
\begin{widetext}

\begin{figure}[t!]
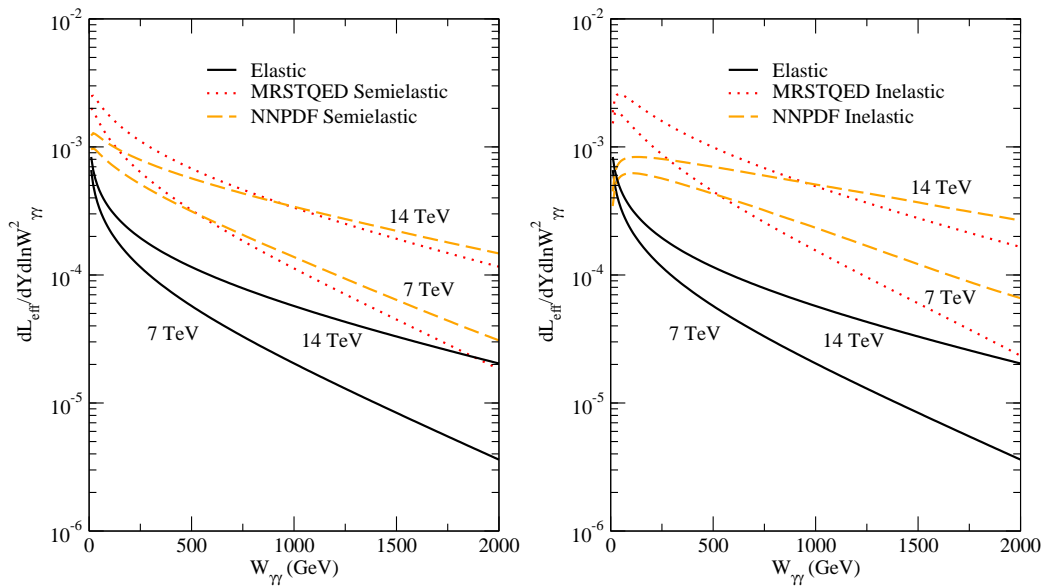

 \includegraphics[scale=0.45]{lumi_invmass_SEMI_clean.eps}
 \includegraphics[scale=0.45]{lumi_invmass_INE_clean.eps}
 \caption{\label{eff_lumi_INE}
(Color online) Comparison between the MRSTQED and NNPDF predictions for effective photon - photon luminosity for semielastic (left) and inelastic (right) processes. The prediction for elastic processes is presented for comparison.
}
\end{figure}

\end{widetext}

\begin{figure}[t!]
\includegraphics[scale=0.35]{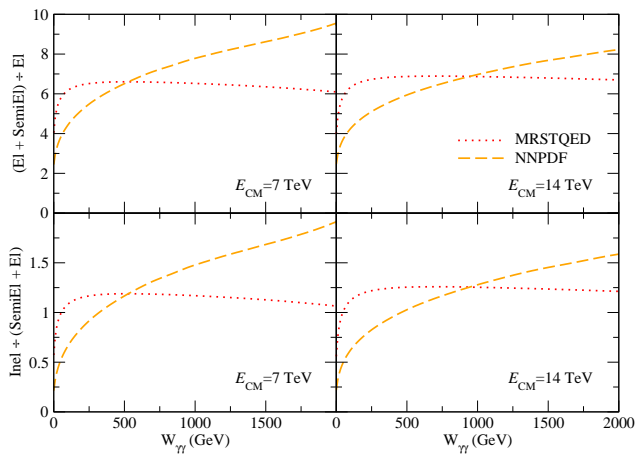}
\caption{\label{ratio_lumi} (Color online) Upper panels: Ratio between the sum of semielastic and elastic luminosities and the elastic luminosity; Lower panels: Ratio between the inelastic luminosity and sum of semielastic and elastic luminosities.}
\end{figure}

Initially lets analyse the impact of the different models for the photon distributions in the effective photon - photon luminosities. We assume that the centre-of-mass energy of the $\gamma \gamma$ system is given by $W_{\gamma\gamma}^2~=~M^{2}$, which implies $x_1 = (W_{\gamma\gamma}/\sqrt{s}) \exp(+Y)$ and $x_2 = (W_{\gamma\gamma}/\sqrt{s}) \exp(-Y)$, and  that $\mu^2 = W_{\gamma\gamma}^2$. In Fig.~\ref{eff_lumi_naive} we assume $Y = 0$ and present the predictions for the  different effective photon - photon luminosities considering the MRSTQED and NNPDF parametrizations for the inelastic photon distributions. Both parametrizations predict that the inelastic contribution is dominant at large  $W_{\gamma\gamma}$, with the behaviour in the region of small $W_{\gamma\gamma}$ being strongly dependent on the model used. It is verified in Fig.~\ref{eff_lumi_INE}, where we compare the semielastic and inelastic predictions from the two parametrizations. We have that the MRSTQED prediction at small $W_{\gamma\gamma}$ is ever larger than the  NNPDF one.  It is clear  that the semielastic and inelastic contributions are computed in a very different way in both parametrizations, leading to distinct behaviours of the effective luminosities in all range of invariant mass. It is important to emphasize the  subtle differences in low mass region. The results with the NNPDF parametrization show a decreasing of the effective luminosity when $W_{\gamma\gamma}$ becomes small, which is not the case for the MRSTQED parametrization. In order to estimate the relative contribution of the semielastic processes with relation to the elastic one, which is important if only one proton is tagged in the final state, in Fig.~\ref{ratio_lumi} (upper panels) we present the ratio between the sum of semielastic and elastic effective luminosities and the elastic one. We obtain that the MRSTQED predictions are almost independent of $W_{\gamma\gamma}$ and $\sqrt{s}$ for $W_{\gamma\gamma} > 150$ GeV. In contrast, the   NNPDF  predictions increases with  $W_{\gamma\gamma}$ and decreases with $\sqrt{s}$. A similar behaviour is obtained for the ratio between the inelastic luminosity and the sum of the semielastic and elastic ones presented in the lower panels of Fig.~\ref{ratio_lumi}.

Lets now estimate the rapidity and invariant mass dependencies of the cross sections for the $\mu^+ \mu^-$ and  $W^+ W^-$ production in two-photon interactions. In Fig. \ref{xsec_mumu_Y} (upper panels) we present our predictions for the rapidity distributions for the  $\mu^+ \mu^-$ production considering the MRSTQED and NNPDF parametrizations, as well as the prediction for elastic processes. We have that the MRSTQED and NNPDF predictions for the rapidity distributions differ significantly. In comparison to the elastic contribution, we have that the semielastic  MRSTQED prediction dominates at central rapidities and the NNPDF one dominates for large values of rapidities, with the inelastic contribution being a factor $\approx 4$ smaller for $Y = 0$ at $\sqrt{s}=$~7~TeV. At $\sqrt{s}=$~14~TeV this factor is also approximately 4 and the semielastic MRSTQED contribution dominates. Such distinct behaviours make the analysis of the rapidity distribution ideal to discriminate between the different models for the inelastic photon distribution. In Fig.\ref{xsec_mumu_Y} (lower panels) we present our predictions for the invariant mass dependencies of the elastic, semielastic and inelastic cross sections. We have that the behaviour observed for the effective photon - photon luminosities is directly reflected in the cross sections, with the NNPDF predictions being larger than the MRSTQED one for large $M$.
We have that the elastic contribution is ever smaller than the semielastic and inelastic contributions, independent of the model used for the inelastic photon distribution. In Fig.~\ref{xsec_wpwm_Y} we present the corresponding distributions for $W^+ W^-$ production. In this case the shape of the rapidity distributions predicted by the MRSTQED and NNPDF parametrizations are similar, differing only in its magnitudes, which are much larger than the prediction for elastic processes at central rapidities. Such dominance also is present in the invariant mass distributions. 

Finally, in Fig.~\ref{ratio_xsec} we present the same ratios presented in Fig.~\ref{ratio_lumi}, but now between  the different cross sections. We obtain that these ratios are almost independent of the final state considered, but strongly dependent on the inelastic photon distribution.   
In particular, considering that a very forward detector is expected to be installed in a short time  in one of the sides of the CMS detector, the determination of the semielastic contribution should be feasible. From the Fig.~\ref{ratio_xsec} (upper panels) we have that the ratio between the sum of elastic and semielastic cross section and the elastic one is almost constant in the MRSTQED case, and strongly increases with $M$ if the NNPDF parametrization is used as input in the calculations. Assuming that elastic contribution is well known, we have that the analysis of this ratio is a direct probe of the inelastic photon distribution. Another important aspect which is distinct in the MRSTQED and NNPDF predictions, is the relative contribution of the inelastic processes show in Fig.~\ref{ratio_xsec} (lower panels), which is predicted to be almost constant in the MRSTQED case, but increases with $M$ in the NNPDF case.  Consequently, the predictions in the region of large invariant masses, where New Physics is expected to be present, are strongly dependent on the inelastic photon parametrization.
In the low-mass region, the predictions are closer. However, in the region of the $WW$ threshold  ($M\approx 160$ GeV), the relative difference at 7~(14) TeV between the MRSTQED and NNPDF predictions is about 2 (16) \% for the first ratio and about 47 (37) \% for the second ratio.  Therefore,  we also have at small invariant masses a non-negligible theoretical uncertainty, directly associated to the description of the photon distribution.

\begin{widetext}

\begin{figure}[t]
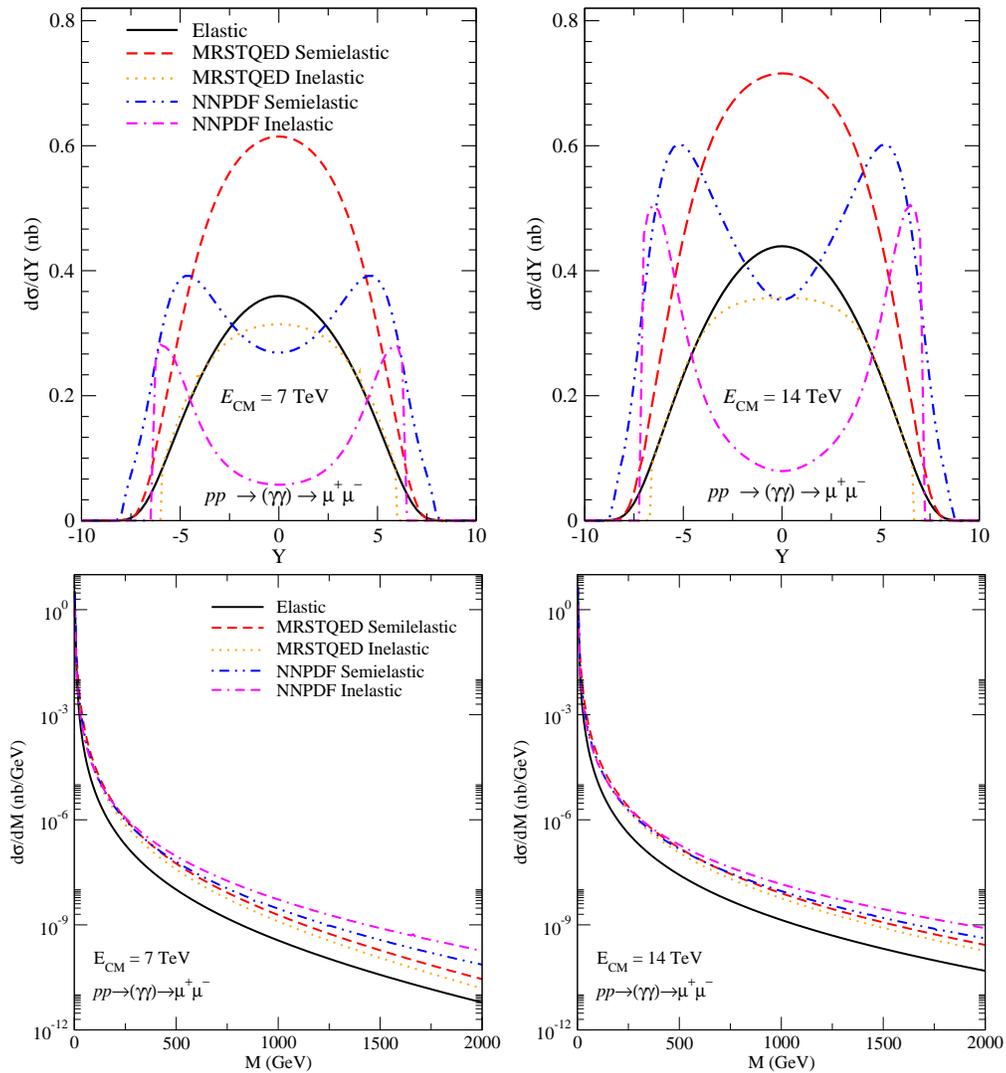

\begin{tabular}{cc}
 \includegraphics[scale=0.45]{mumu_Y_7TeV_clean.eps} &
 \includegraphics[scale=0.45]{mumu_Y_14TeV_clean.eps} \\
 \includegraphics[scale=0.4]{mumu_M_7TeV_clean.eps} &
 \includegraphics[scale=0.4]{mumu_M_14TeV_clean.eps}
 \end{tabular}
  \caption{\label{xsec_mumu_Y} Rapidity and invariant mass distributions for the $\mu^+ \mu^-$ production considering the MRSTQED and NNPDF parametrizations. The  prediction for elastic processes  is presented for comparison.
}

\end{figure}


\begin{figure}[t]
\begin{tabular}{cc}
 \includegraphics[scale=0.45]{wpwm_Y_7TeV_clean.eps} &
 \includegraphics[scale=0.45]{wpwm_Y_14TeV_clean.eps} \\
 \includegraphics[scale=0.4]{wpwm_M_7TeV_clean.eps} &
 \includegraphics[scale=0.4]{wpwm_M_14TeV_clean.eps}
 \end{tabular}
  \caption{\label{xsec_wpwm_Y}  Rapidity and invariant mass distributions for the $W^{+}W^{-}$ production considering the MRSTQED and NNPDF parametrizations. The  prediction for elastic processes  is presented for comparison. 
}

\end{figure}


\begin{figure}[t]

 \includegraphics[scale=0.5]{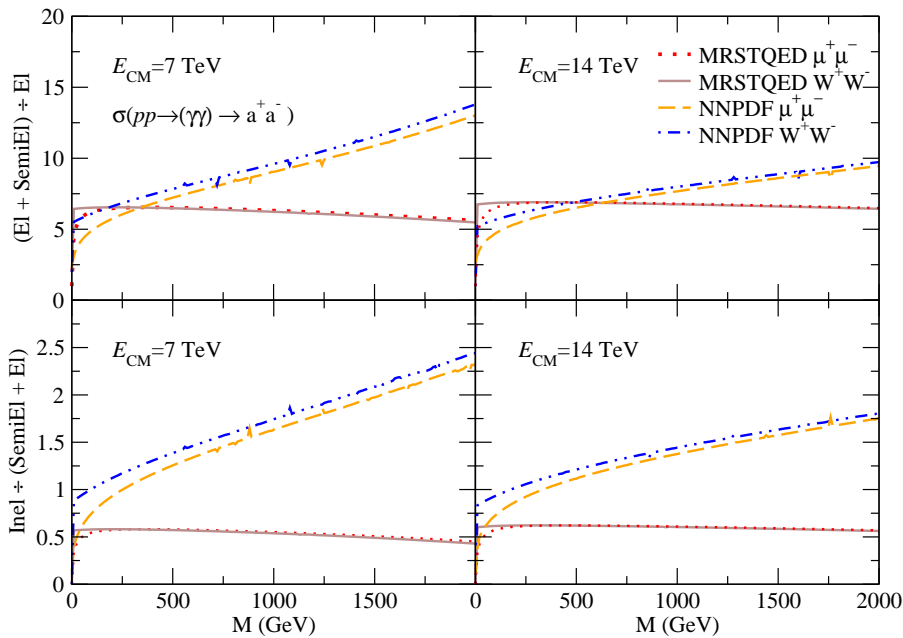}
 \caption{\label{ratio_xsec} (Color online) Upper panels: Ratio between the sum of semielastic and elastic cross sections and the elastic cross section; Lower panels: Ratio between the inelastic cross section and sum of semielastic and elastic cross sections.}
\end{figure}

\end{widetext}

\section{Summary} 
\label{sec:sum}

In this paper we have investigated the contribution of inelastic, semielastic and elastic processes for the $W^+W^-$ and $\mu^+\mu^-$ production by two-photon interactions in $pp$ collisions considering different models for the inelastic photon distribution. 
We demonstrated that these distinct models implies very distinct behaviours for the effective photon - photon luminosities and differential cross sections. It implies that the use of the two-photon particle production mechanism to search rare events will be not a easy task before the determination of the correct description of the inelastic photon distribution. In particular, since in a first moment only one of the very forward detectors of CT-PPS will be installed, which would not eliminate the semielastic processes. Our results indicate that the analysis of the rapidity distribution for $\mu^+\mu^-$ production
can be a discriminator among the possible models for the photon distribution. In particular, the determination of the distribution for central rapidities already differentiate the distinct models. Moreover, the analysis of the invariant mass dependence of the ratio between the sum of semielastic and elastic cross sections and the elastic one also can be used to determinate the inelastic photon distribution. Finally, we believe that the analyses carried out with the data of the two-photon production of pairs, in conjunction with the exclusive vector meson production discussed in Ref.~\cite{vicgus}, will allow us to precisely determine the adequate description of the photon distribution, which still is an important open question in High Energy Physics.

\section*{Acknowledgements}
This work has been supported by CNPq, CAPES and FAPERGS, Brazil.


\begin{thebibliography}{}

\bibitem{upc}
 G. Baur, K. Hencken, D. Trautmann, S. Sadovsky, Y. Kharlov, Phys.
Rep. {\bf 364}, 359 (2002); 
V.~P.~Goncalves and M.~V.~T.~Machado,
Mod. Phys. Lett. A {\bf 19}, 2525  (2004); 
 C.~A. Bertulani, S.~R.~Klein and J.~Nystrand, Ann. Rev. Nucl. Part. Sci. {\bf 55}, 
271 (2005);
 K.~Hencken {\it et al.},
  Phys.\ Rept.\  {\bf 458}, 1 (2008).


\bibitem{epa} 
  V.~M.~Budnev, I.~F.~Ginzburg, G.~V.~Meledin and V.~G.~Serbo,
  Phys.\ Rept.\  {\bf 15}, 181 (1975).

\bibitem{Terazawa} 
  H.~Terazawa,
  Rev.\ Mod.\ Phys.\  {\bf 45}, 615 (1973).

\bibitem{Baurjpg} 
  G.~Baur, K.~Hencken and D.~Trautmann,
  J.\ Phys.\ G {\bf 24}, 1657 (1998)
  

\bibitem{Baurdileptons} 
  G.~Baur, K.~Hencken and D.~Trautmann,
  Phys.\ Rept.\  {\bf 453}, 1 (2007)

\bibitem{Chernyak} 
  V.~L.~Chernyak and S.~I.~Eidelman,
  Prog.\ Part.\ Nucl.\ Phys.\  {\bf 80}, 1 (2014)

\bibitem{Chatrchyan:2011ci} 
  S.~Chatrchyan {\it et al.}  [CMS Collaboration],
  JHEP {\bf 01}, 052 (2012)
  
\bibitem{Chatrchyan:2012tv} 
  S.~Chatrchyan {\it et al.}  [CMS Collaboration],
  JHEP {\bf 11}, 080 (2012)  

\bibitem{Chatrchyan:2013foa} 
  S.~Chatrchyan {\it et al.}  [CMS Collaboration],
  JHEP {\bf 07}, 116 (2013)
  
  
\bibitem{newphysics}
J.~de Favereau de Jeneret, V.~Lemaitre, Y.~Liu, S.~Ovyn, T.~Pierzchala, K.~Piotrzkowski, X.~Rouby and N.~Schul {\it et al.},
  arXiv:0908.2020 [hep-ph]; 
S.~Atag, S.~C.~Inan and I.~Sahin,
  JHEP {\bf 1009}, 042 (2010); S.~C.~Inan,
  Phys.\ Rev.\ D {\bf 81}, 115002 (2010); E.~Chapon, C.~Royon and O.~Kepka,
  Phys.\ Rev.\ D {\bf 81}, 074003 (2010); R.~S.~Gupta,
  Phys.\ Rev.\ D {\bf 85}, 014006 (2012);  H.~Sun,
  Eur.\ Phys.\ J.\ C {\bf 74}, no. 8, 2977 (2014);
    P.~Lebiedowicz, R.~Pasechnik and A.~Szczurek,
  Nucl.\ Phys.\ B {\bf 881}, 288 (2014);
  V.~P.~Goncalves, W.~K.~Sauter and M.~Thiel,
  Phys.\ Rev.\ D {\bf 89}, 076003 (2014); P.~Lebiedowicz and A.~Szczurek,
  Phys.\ Rev.\ D {\bf 91}, no. 9, 095008 (2015); 
  V.~P.~Goncalves and W.~K.~Sauter,
  Phys.\ Rev.\ D {\bf 91}, 035004 (2015); S.~Fichet, G.~von Gersdorff, B.~Lenzi, C.~Royon and M.~Saimpert,
  JHEP {\bf 1502}, 165 (2015).

\bibitem{Fermi} 
  E.~Fermi,
  Z.\ Phys.\  {\bf 29}, 315 (1924).

\bibitem{Williams} 
  E.~J.~Williams,
  Phys.\ Rev.\  {\bf 45}, 729 (1934).


\bibitem{Weizsacker} 
  C.~F.~von Weizsacker,
  Z.\ Phys.\  {\bf 88}, 612 (1934).
  

  
  
  

\bibitem{kniehl} 
  B.~A.~Kniehl,
  Phys.\ Lett.\ B {\bf 254}, 267 (1991).

\bibitem{rujula} 
  A.~De Rujula and W.~Vogelsang,
  Phys.\ Lett.\ B {\bf 451}, 437 (1999) 
  
\bibitem{drees_godbole} 
  M.~Drees, R.~M.~Godbole, M.~Nowakowski and S.~D.~Rindani,
  Phys.\ Rev.\ D {\bf 50}, 2335 (1994)  

\bibitem{pisano} 
  M.~Gluck, C.~Pisano and E.~Reya,
  Phys.\ Lett.\ B {\bf 540}, 75 (2002); C.~Pisano,
  hep-ph/0512306.

\bibitem{mrstqed} 
  A.~D.~Martin, R.~G.~Roberts, W.~J.~Stirling and R.~S.~Thorne,
  Eur.\ Phys.\ J.\ C {\bf 39}, 155 (2005)

\bibitem{nnpdf} 
  R.~D.~Ball {\it et al.}  [NNPDF Collaboration],
  Nucl.\ Phys.\ B {\bf 877}, 290 (2013)
 
 \bibitem{cteqqed} 
  C.~Schmidt, J.~Pumplin, D.~Stump and C.-P.~Yuan,
  PoS DIS {\bf 2014}, 054 (2014).

\bibitem{martin_ryskin} 
  A.~D.~Martin and M.~G.~Ryskin,
  Eur.\ Phys.\ J.\ C {\bf 74}, 3040 (2014) 




\bibitem{vicgus} 
  V.~P.~Goncalves and G.~G.~da Silveira,
  Phys.\ Rev.\ D {\bf 91}, no. 5, 054013 (2015)

\bibitem{antoni_gus} 
  G.~G.~da Silveira, L.~Forthomme, K.~Piotrzkowski, W.~Schäfer and A.~Szczurek,
  JHEP {\bf 1502}, 159 (2015)
 
\bibitem{antoni_royon} 
  M.~Luszczak, A.~Szczurek and C.~Royon,
  JHEP {\bf 1502}, 098 (2015)  


\bibitem{Albrow:2008pn} 
  M.~G.~Albrow {\it et al.}  [FP420 R and D Collaborations],
  JINST {\bf 4}, T10001 (2009)



\bibitem{ctpps}
  The CMS and TOTEM Collaborations, CMS-TOTEM Precision Proton Spectrometer Technical Design Report, \url{http://cds.cern.ch/record/1753795}.	


\bibitem{d'Enterria:2013yra} 
  D.~d'Enterria and G.~G.~da Silveira,
  Phys.\ Rev.\ Lett.\  {\bf 111}, 080405 (2013)


\bibitem{KMR} 
  V.~A.~Khoze, A.~D.~Martin and M.~G.~Ryskin,
  Eur.\ Phys.\ J.\ C {\bf 23}, 311 (2002)
  
\bibitem{inel}
J.~Ohnemus, T.~F.~Walsh and P.~M.~Zerwas,
  Phys.\ Lett.\ B {\bf 328}, 369 (1994); G.~Bhattacharya, P.~Kalyniak and K.~A.~Peterson,
  Phys.\ Rev.\ D {\bf 53}, 2371 (1996);    D.~d'Enterria and J.~P.~Lansberg,
  Phys.\ Rev.\ D {\bf 81}, 014004 (2010) 
  
  
\bibitem{vicwerdaniel}  
 V.~P.~Goncalves, D.~T.~da Silva and W.~K.~Sauter,
  Phys.\ Rev.\ C {\bf 87}, 028201 (2013)

\bibitem{Boonekamp:2007iu} 
  M.~Boonekamp, C.~Royon, J.~Cammin and R.~B.~Peschanski,
  Phys.\ Lett.\ B {\bf 654}, 104 (2007)


\end{thebibliography}
\end{document}